\documentclass[longauth]{aa}
\usepackage{graphicx}
\usepackage{txfonts}
\usepackage{xcolor}
\usepackage{hyperref}
\hypersetup{colorlinks, linkcolor=blue, citecolor=blue, urlcolor=blue}
\usepackage{amsmath}
\usepackage{ulem}
\usepackage{soul}
\usepackage{enumitem}

\newcommand{\pivec}{\mbox{\boldmath $\pi$}}
\newcommand{\muvec}{\mbox{\boldmath $\mu$}}

\newcommand{\thetavec}{\mbox{\boldmath $\theta$}}
\newcommand{\te}{t_{\rm E}}
\newcommand{\thetae}{\theta_{\rm E}}

\newcommand{\pie}{\pi_{\rm E}}

\newcommand{\dl}{D_{\rm L}}

\definecolor{brown}{rgb}{0.59, 0.29, 0.0}
\definecolor{darkgreen}{rgb}{0.0, 0.42, 0.24}
\definecolor{darkblue}{rgb}{0.01, 0.31, 0.59}
\definecolor{darkblue}{rgb}{0.0, 0.25, 0.42}
\definecolor{blue}{rgb}{0.0,0.0,1.0}
\definecolor{green}{rgb}{0.0,1.0,0.0}

\def\eqalign#1{\null\,\vcenter{\openup\jot
        \ialign{\strut\hfil$\displaystyle{##}$&$
        \displaystyle{{}##}$\hfil \crcr#1\crcr}}\,}

\begin{document}

\title{KMT-2021-BLG-0912Lb: A microlensing super Earth around a K-type star}

\author{
     Cheongho~Han\inst{01} 
\and Ian~A.~Bond\inst{02} 
\and Jennifer~C.~Yee\inst{03}   
\and Weicheng~Zang\inst{04}
\\
(Leading authors)\\
     Michael~D.~Albrow\inst{05}   
\and Sun-Ju~Chung\inst{06}      
\and Andrew~P.~Gould\inst{07,08}      
\and Kyu-Ha~Hwang\inst{06} 
\and Youn~Kil~Jung\inst{06} 
\and Doeon~Kim\inst{01}
\and Chung-Uk~Lee\inst{06} 
\and Yoon-Hyun~Ryu\inst{06} 
\and In-Gu~Shin\inst{06} 
\and Yossi~Shvartzvald\inst{09}   
\and Sang-Mok~Cha\inst{06,10} 
\and Dong-Jin~Kim\inst{06} 
\and Seung-Lee~Kim\inst{06,11} 
\and Dong-Joo~Lee\inst{06} 
\and Yongseok~Lee\inst{06} 
\and Byeong-Gon~Park\inst{06,11} 
\and Richard~W.~Pogge\inst{08}
\\
(The KMTNet Collaboration),\\
     Fumio~Abe\inst{12}
\and Richard~Barry\inst{13}
\and David~P.~Bennett\inst{13,14}
\and Aparna~Bhattacharya\inst{13,14}
\and Yuki~Hirao\inst{15}
\and Hirosane~Fujii\inst{15}
\and Akihiko~Fukui\inst{16,17}
\and Yoshitaka~Itow\inst{12}
\and Rintaro~Kirikawa\inst{15}
\and Iona~Kondo\inst{15}
\and Naoki~Koshimoto\inst{18,19}
\and Yutaka~Matsubara\inst{12}
\and Sho~Matsumoto\inst{15}
\and Yasushi~Muraki\inst{12}
\and Shota~Miyazaki\inst{15}
\and Cl\'ement~Ran\inst{13}
\and Arisa~Okamura\inst{15}
\and Nicholas~J.~Rattenbury\inst{20}
\and Yuki~Satoh\inst{15}
\and Takahiro~Sumi\inst{15}
\and Daisuke~Suzuki\inst{21}
\and Stela~Ishitani~Silva\inst{13,22}
\and Taiga~Toda\inst{15}
\and Paul~J.~Tristram\inst{23}
\and Hibiki~Yama\inst{15}
\and Atsunori~Yonehara\inst{15}
\\
(The MOA Collaboration),\\
     Tony~Cooper\inst{24}
\and Plamen~Dimitrov\inst{24}
\and Subo~Dong\inst{25}
\and John~Drummond\inst{26,27}
\and Jonathan~Green\inst{24}
\and Steve~Hennerley\inst{24}
\and Zhuokai~Liu\inst{25,28}
\and Shude~Mao\inst{04,29}
\and Dan~Maoz\inst{30}
\and Matthew~Penny\inst{31}
\and Hongjing~Yang\inst{04}
\\
(LCOGT \& $\mu$FUN Follow-up Team),\\ 
}

\institute{
     Department of Physics, Chungbuk National University, Cheongju 28644, Republic of Korea  \\ \email{cheongho@astroph.chungbuk.ac.kr}                  
\and Institute of Natural and Mathematical Sciences, Massey University, Auckland 0745, New Zealand                                                       
\and Center for Astrophysics $|$ Harvard \& Smithsonian 60 Garden St., Cambridge, MA 02138, USA                                                          
\and Department of Astronomy, Tsinghua University, Beijing 100084, China                                                                                 
\and University of Canterbury, Department of Physics and Astronomy, Private Bag 4800, Christchurch 8020, New Zealand                                     
\and Korea Astronomy and Space Science Institute, Daejon 34055, Republic of Korea                                                                        
\and Max Planck Institute for Astronomy, K\"onigstuhl 17, D-69117 Heidelberg, Germany                                                                    
\and Department of Astronomy, The Ohio State University, 140 W. 18th Ave., Columbus, OH 43210, USA                                                       
\and Department of Particle Physics and Astrophysics, Weizmann Institute of Science, Rehovot 76100, Israel                                               
\and School of Space Research, Kyung Hee University, Yongin, Kyeonggi 17104, Republic of Korea                                                           
\and Korea University of Science and Technology, 217 Gajeong-ro, Yuseong-gu, Daejeon, 34113, Republic of Korea                                           
\and Institute for Space-Earth Environmental Research, Nagoya University, Nagoya 464-8601, Japan                                                         
\and Code 667, NASA Goddard Space Flight Center, Greenbelt, MD20771, USA                                                                                 
\and Department of Astronomy, University of Maryland, College Park, MD 2074                                                                              
\and Department of Earth and Space Science, Graduate School of Science, Osaka University, Toyonaka, Osaka 560-0043, Japan                                
\and Department of Earth and Planetary Science, Graduate School of Science, The University of Tokyo, 7-3-1 Hongo, Bunkyo-ku, Tokyo 113-0033, Japan       
\and Instituto de Astrof\'sica de Canarias, V\'ia L\'actea s/n, E-38205 La Laguna, Tenerife, Spain                                                       
\and Department of Astronomy, Graduate School of Science, The University of Tokyo, 7-3-1 Hongo, Bunkyo-ku, Tokyo 113-0033, Japan                         
\and National Astronomical Observatory of Japan, 2-21-1 Osawa, Mitaka, Tokyo 181-8588, Japan                                                             
\and Department of Physics, University of Auckland, Private Bag 92019, Auckland, New Zealand                                                             
\and Institute of Space and Astronautical Science, Japan Aerospace Exploration Agency, 3-1-1 Yoshinodai, Chuo, Sagamihara, Kanagawa, 252-5210, Japan     
\and Department of Physics, The Catholic University of America, Washington, DC 20064, USA                                                                
\and University of Canterbury Mt. John Observatory, P.O. Box 56, Lake Tekapo 8770, New Zealand                                                           
\and Kumeu Observatory, Kumeu, New Zealand                                                                                                               
\and Kavli Institute for Astronomy and Astrophysics, Peking University, Yi He Yuan Road 5, Hai Dian District, Beijing 100871, China                      
\and Possum Observatory, Patutahi, New Zealand                                                                                                           
\and Centre for Astrophysics, University of Southern Queensland, Toowoomba, Queensland 4350, Australia                                                   
\and Department of Astronomy, School of Physics, Peking University, Yi He Yuan Road 5, Hai Dian District, Beijing 100871, China                          
\and National Astronomical Observatories, Chinese Academy of Sciences, Beijing 100101, China                                                             
\and School of Physics and Astronomy, Tel-Aviv University, Tel-Aviv 6997801, Israel                                                                      
\and Department of Physics and Astronomy, Louisiana State University, Baton Rouge, LA 70803 USA                                                          
}
\date{Received ; accepted}

\abstract
{}
{
The light curve of the microlensing event KMT-2021-BLG-0912 exhibits a very short anomaly relative 
to a single-lens single-source form. We investigate the light curve for the purpose of identifying 
the origin of the anomaly.
}
{
We model the light curve under various interpretations. From this, we find four solutions, in which 
three solutions are found under the assumption that the lens is composed of two masses (2L1S models), 
and the other solution is found under the assumption that the source is comprised of a binary-star 
system (1L2S model).  The 1L2S model is ruled out based on the contradiction that the faint source 
companion is bigger than its primary, and one of the 2L1S solutions is excluded from the combination 
of the relatively worse fit, blending constraint, and lower overall probability,  leaving two surviving 
solutions with the planet/host mass ratios of $q\sim 2.8\times 10^{-5}$ and $\sim 1.1\times 10^{-5}$.  
A subtle central deviation supports the possibility of a tertiary lens component, either a binary 
companion to the host with a very large or small separation or a second planet lying near the Einstein 
ring, but it is difficult to claim a secure detection due to the marginal fit improvement, lack of 
consistency among different data sets, and difficulty in uniquely specifying the nature of the tertiary 
component.
}
{
With the observables of the event, it is estimated that the masses of the planet and host are
$\sim (6.9~M_\oplus, 0.75~M_\odot)$ according to one solution and $\sim (2.8~M_\oplus, 0.80~M_\odot)$ 
according to the other solution, indicating that the planet is a super Earth around a K-type star, 
regardless of the solution.  The fact that 16, including the one reported in this work, out of 19 
microlensing planets with $M \lesssim  10~M_\oplus$ were detected during the last 6 years well 
demonstrates the importance of high-cadence global surveys in detecting very low-mass planets.  
}
{}


\maketitle

\section{Introduction}\label{sec:one}

The current microlensing surveys differentiate themselves from those of the previous era by their 
greatly increased observational cadence.  High-cadence surveys with multiple observations per night 
became possible with the combination of the instrumental upgrade of existing surveys and the inauguration 
of a new survey employing globally distributed multiple telescopes equipped with cameras yielding a very 
large field of view (FOV). The Microlensing Observations in Astrophysics survey \citep[MOA:][]{Bond2001}, 
which originally started its first phase experiment with the use of a 0.61~m telescope and a camera having 
a 1.3~deg$^2$ FOV, entered into its second phase by employing a 1.8~m telescope and replacing its old 
camera with a new one having a 2.2~deg$^2$ FOV. The Optical Gravitational Lensing Experiment (OGLE),  
which started its first-phase experiment with a 1.0~m telescope, is now in its fourth phase 
\citep[OGLE-IV:][]{Udalski2015} using a 1.3~m telescope mounted with a mosaic camera yielding a 1.4~deg$^2$ 
FOV.  The Korea Microlensing Telescope Network \citep[KMTNet:][]{Kim2016} started its full operation in 
2016 using its three 1.6~m telescopes, each of which is equipped with a camera providing a 4~deg$^2$ FOV. 
With the instrumental upgrade of the existing experiments and the advent of a new survey, the cadence of 
the current surveys reaches down to $< 8$~minutes, which is two orders of magnitude higher than that of
earlier experiments.

The increased survey area combined with the increased observational cadence resulted in the dramatic 
rise in the detection rate of lensing events. The annual event rate, which was of the order of dozens 
in the 1990s and hundreds in the 2000s, has soared to more than 3000 from the combination of the three 
lensing surveys. The increase of the total event rate is accompanied by the rapid increase in the 
number of microlensing planet detections, and about 20 planetary systems are annually being reported.

\begin{figure}[t]
\includegraphics[width=\columnwidth]{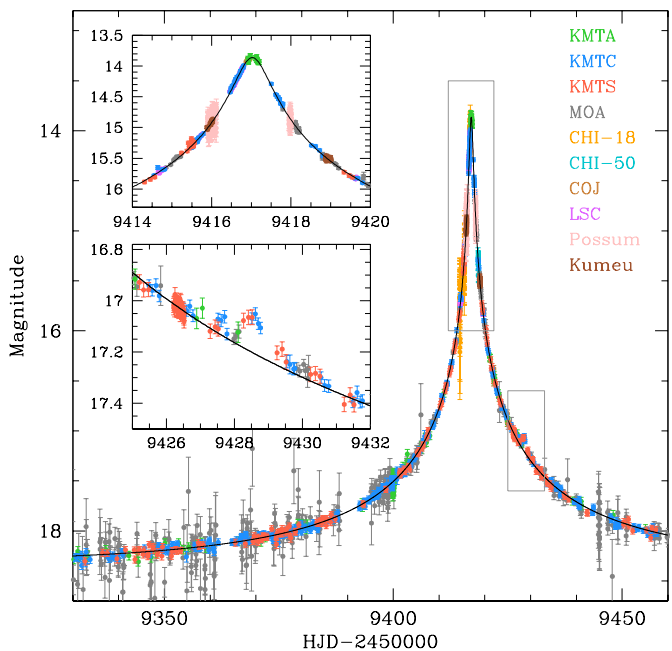}
\caption{
Light curve of KMT-2021-BLG-0912. The insets show the zoom-in views of the peak and anomaly 
regions. The curve drawn over data points is a model based on the single-lens single-source
interpretation.
}
\label{fig:one}
\end{figure}

Another important achievement of high-cadence global surveys is that they expand the channel of 
planet detections.  In general, microlensing signals of planets appear as short-term anomalies 
on the lensing light curves produced by hosts. The duration of the planetary signal becomes shorter 
in proportion to the square root of the planet-to-host mass ratio $q$, that is, $t_{\rm planet} 
\sim q^{1/2}\te$, where $\te$ represents the event time scale.  Then the duration is several days 
for giant planets with $q\sim (O)10^{-3}$ and less than one day\footnote{ In practice, longer 
durations may be observed for oblique source trajectories, e.g., KMT-2020-BLG-0414 \citep{Zang2021} 
and the planet presented here.} for planets with $q\lesssim (O)10^{-4}$.  As a result, it was difficult 
to detect planets with very low mass ratios from observations conducted in a survey-only mode with the 
$\sim 1$~day cadence of the earlier generation surveys.  To complement the inadequate cadence of the 
previous surveys, planetary lensing experiments had been carried out in a survey+followup mode, in 
which survey experiments focused mostly on event detections and followup experiments coordinated 
dense global coverage of the events detected by the survey experiments, for example, OGLE-2005-BLG-071Lb 
\citep{Udalski2005, Dong2009} and OGLE-2005-BLG-390Lb \citep{Beaulieu2006}.  Dense followup observations 
require alerts of events with ongoing anomalies, but such observations were difficult for events with 
very short planetary signals due to the difficulty of predicting the time of the anomaly and immediately 
conducting followup observations.  As a result, observations in the survey+followup mode were conducted 
most successfully for a special type of events with very high magnifications, for which not only the 
time of the peak could be predicted with a reasonable precision but also the chance of planetary anomaly 
near the peak was high: the ``central magnification channel'' \citep{Griest1998}.  High-cadence surveys 
with globally distributed telescopes dispense with the need for alerts for dense coverage of planetary 
signals, and thus they expand the channel of planet detections to all events regardless of lensing 
magnifications, for example, OGLE-2018-BLG-0567Lb and OGLE-2018-BLG-0962Lb \citep{Jung2021}.

In this work, we report the discovery of a very low mass-ratio planet found from the analysis of a 
lensing event detected in the 2021 season.  Despite the short duration and weakness, the planetary 
signal, which occurred well after the high-magnification peak, was clearly detected by the dense 
coverage of the lensing surveys.  We test various interpretations to confirm the planetary origin 
of the anomaly.

We present the analysis according to the following organization. In Sect.~\ref{sec:two}, we give an 
explanation for the observations of the lensing event and the acquired data.  In Sect.~\ref{sec:three}, 
we describe the characteristics of the light-curve anomaly and depict various models tested to explain 
the origin of the anomaly, including binary-lens and binary-source models. We detail the procedures of 
the modeling and present results. We additionally test the possibility that the lens is composed of three 
masses. The procedures of characterizing the source star and estimating the angular Einstein radius are 
depicted in Sect.~\ref{sec:four}. We estimate the physical parameters of the planetary system in 
Sect.~\ref{sec:five}, and we summarize results and conclude in Sect.~\ref{sec:six}.

\section{Observations and data}\label{sec:two}

The planet we report in this work was detected from the observations of the lensing event
KMT-2021-BLG-0912/MOA-2021-BLG-233. The source of the event is located in the Galactic bulge
field with the equatorial coordinates (RA, Decl.)$_{\rm J2000}$ $ = $ (17:59:17.08, -31:59:51.50), 
corresponding to the Galactic coordinates $(l, b)= (-1^\circ\hskip-2pt .142, -4^\circ\hskip-2pt .091)$. 
The apparent baseline magnitude of the source is $I_{\rm base}\sim 18.31$ according to the KMTNet scale.

\begin{table}[t]
\small
\caption{Followup data\label{table:one}}
\begin{tabular*}{\columnwidth}{@{\extracolsep{\fill}}lll}
\hline\hline
\multicolumn{1}{c}{Telescopes}             &
\multicolumn{1}{c}{Epoch (HJD$^\prime$)}   &
\multicolumn{1}{c}{Aperture}               \\
\hline
LCO COJ                &  9416, 9421, 9424              &  1.0~m    \\    
LCO LSC                &  9414, 9415, 9416, 9419        &  1.0~m    \\
CHI-18                 &  9414, 9415, 9416              &  0.18~m   \\
CHI-50                 &  9416, 9418                    &  0.5~m    \\
$\mu$FUN Possum        &  9415, 9416, 9417, 9418        &  0.4~m    \\
$\mu$FUN Kumeu         &  9415, 9416, 9418              &  0.4~m    \\
\hline
\end{tabular*}
\tablefoot{ ${\rm HJD}^\prime\equiv {\rm HJD}-2450000$.  }
\end{table}

The magnification of the source flux caused by lensing was first found by the KMTNet survey on
2021-05-19 (${\rm HJD}^\prime\equiv {\rm HJD}-2450000\sim 9353$) at its early stage of the lensing 
magnification.  The event was independently found by the MOA survey 51 days after the KMTNet 
discovery, and it was labeled as MOA-2021-BLG-233. 
Hereafter, we refer to the event as KMT-2021-BLG-0912 following the convention of designating a 
lensing event by the ID of the survey that first discovered the event. Data from the KMTNet survey 
were acquired using the three 1.6~m telescopes located at the Siding Spring Observatory (SSO) in 
Australia (KMTA), Cerro Tololo Inter-American Observatory (CTIO) in Chile (KMTC), and South African 
Astronomical Observatory (SAAO) in South Africa (KMTS). The MOA survey utilized the 1.8~m telescope 
of the Mt.~John Observatory in New Zealand.  Images were mainly acquired in the $I$ band for the KMTNet 
survey and in the customized MOA-$R$ band for the MOA survey. For both surveys, a fraction of images 
were obtained in the $V$ band for the measurement of the source color.  The event was also in the field 
of the OGLE survey.  However, no OGLE observation was done because the OGLE telescope was shut down in 
the 2020 and 2021 seasons due to the Covid-19 pandemic.

\begin{figure}[t]
\includegraphics[width=\columnwidth]{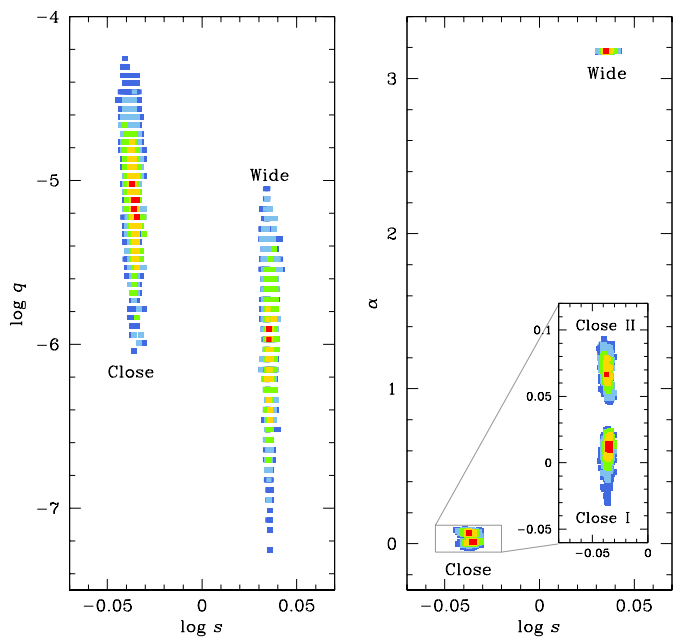}
\caption{
$\Delta\chi^2$ map on the $\log s$--$\log q$ (left panel) and $\log s$--$\alpha$ (right panel) planes 
obtained from the 2L1S modeling. The maps shows three locals: close~I $(\log s, \log q, \alpha)\sim 
(-0.04, -5.1, 0.01)$, close~II $\sim (-0.04, -5.1, 0.07)$, and wide $\sim (0.04, -5.9, 3.2)$ solutions. 
The inset in the right panel shows the zoom-in view of around the close~I and close~II solutions on the 
$\log~s$--$\alpha$ (right panel) plane. The color coding is set to represents points with 
$\leq 1n\sigma$ (red), 
$\leq 2n\sigma$ (yellow), 
$\leq 3n\sigma$ (green), 
$\leq 4n\sigma$ (cyan), and 
$\leq 5n\sigma$ (blue), where $n=2$.
}
\label{fig:two}
\end{figure}

The event was additionally covered by followup observations.  At 2021 July 16, UT 21:30 
(${\rm HJD}^\prime\sim 9412.4$), the KMTNet HighmagFinder system (Yang et al., in preparation) 
predicted that the event would reach a very high magnification based on the KMTNet data on the 
rising side of the light curve.  Because the chance to detect planetary signals was high near 
the peak, the KMTNet group issued an alert for followup observations. In response to this alert, 
follow-up observations were conducted around the peak of the light curve using multiple telescopes 
that were globally distributed, including the two 1.0~m telescopes, one located at SSO (COJ) and 
the other at CTIO (LSC), of the Las Cumbres Observatory (LCO) global network, the 0.18~m and 0.5~m 
Newtonian telescopes located at El Sauce Observatory in Chile (CHI-18 and CHI-50), the two telescopes 
of the Microlensing Follow-Up Network ($\mu$FUN) group (Possum 0.4~m and Kumeu 0.4~m) located in New 
Zealand.  In Table~\ref{table:one}, we list the follow-up telescopes and the epochs of observations. 
Furthermore, the KMTNet survey increased its observational cadence by employing the ``auto-followup'' 
system, which began its operation in the 2021 season for dense coverage of high-magnification events.  
However, even with the dense coverage by the combined data from the survey and follow-up observations, 
there is no significant anomaly over the peak of the light curve, although these observations do help 
constrain the finite source effect.  The zoom-in view around the peak region is shown in the upper 
inset of Figure~\ref{fig:one}. In Sect.~\ref{sec:three-three}, we will give an in-depth discussion 
of a possible central anomaly.

A short-term anomaly occurring much later (${\rm HJD}^\prime \sim 9428.5$) was noticed by C.~Han from  
a revisit of the event on August 19 (${\rm HJD}^\prime \sim 9437$), when the source brightness was 
declining. The zoom-in view around the anomaly is presented in the lower inset of Figure~\ref{fig:one}. 
The anomaly, which was covered mostly by the KMTC and KMTS data sets, lasted for $\sim 1$~day centered 
at ${\rm HJD}^\prime\sim 9428.5$.  At the peak of the anomaly, the data points deviate from the 1L1S 
model by $\sim 0.2$~mag. After the anomaly, the event gradually returned to the baseline.

\begin{figure}[t]
\includegraphics[width=\columnwidth]{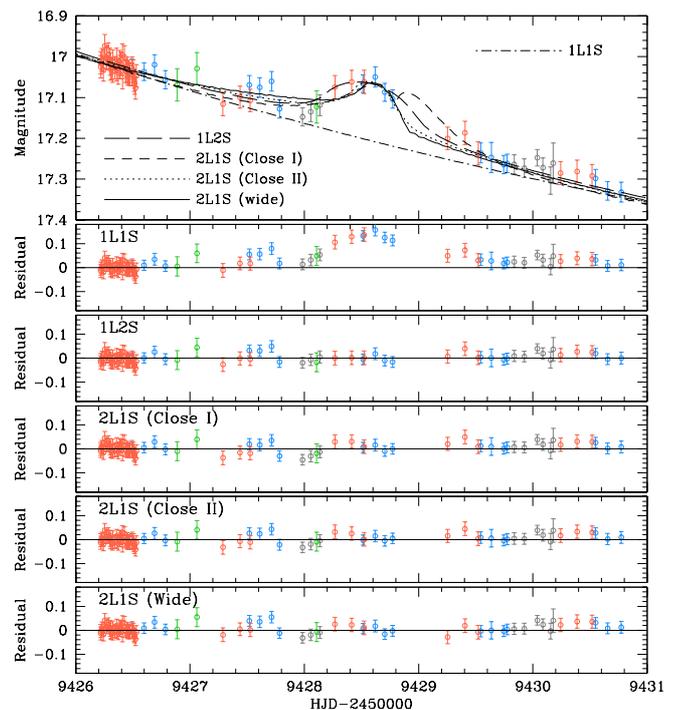}
\caption{
Models curves and residuals of the five tested solutions, including the 1L1S, 1L2S, three 2L1S 
(close~I, close~II, and wide) solutions.
}
\label{fig:three}
\end{figure}

The data obtained by the KMTNet and MOA surveys were processed utilizing the pipelines of the 
individual groups: \citet{Albrow2009} for the KMTNet survey and \citet{Bond2001} for the MOA 
survey. These pipelines are developed based on the difference image technique \citep{Tomaney1996, 
Alard1998}.  The followup data from the LCO, CHI-18, and CHI-50 observations were reduced using 
the ISIS pipeline \citep{Alard2000, Zang2018}, and those from the $\mu$FUN observations were 
reduced using the DoPHOT code \citep{Schechter1993}. For the individual data sets, the error 
bars estimated from the photometry pipelines were readjusted to take into consideration the 
scatter of data and to make the $\chi^2$ per degree of freedom for each data set unity following 
the routine described in \citet{Yee2012}.

Figure~\ref{fig:one} shows the light curve of KMT-2021-BLG-0912. The curve drawn over the data points 
is a 1L1S model with lensing parameters $(t_0, u_0, \te)\sim (9417.02, 6.0 \times 10^{-3}, 70$~days). 
Here $t_0$ is the time, expressed in ${\rm HJD}^\prime$, of the closest lens-source approach and $u_0$ 
is the separation at $t_0$.  The two small boxes drawn over the light curve indicate the regions around 
the peak and the anomaly.  The magnification at the peak was $A_{\rm peak} \sim 1/u_0\sim 170$.

\section{Interpreting the event}\label{sec:three}

The light curve of the lensing event KMT-2021-BLG-0912 is characterized by a short-lasting anomaly 
on the otherwise smooth and symmetric 1L1S curve. This is a characteristic pattern of an event 
produced by a lens with a very low-mass companion \citep{Gould1992b}. Therefore, we first model the 
light curve under a binary-lens single-source (2L1S) interpretation.  We additionally test a single-lens 
binary-source (1L2S) interpretation because it is known that a subset of 1L2S events can generate 
short-term anomalies in lensing light curves \citep{Gaudi1998a, Gaudi2004}.

Considering an additional lens or source component in lensing modeling requires one to include
extra parameters in addition to the 1L1S lensing parameters, that is, $(t_0, u_0, \te)$. For a 
2L1S model, these extra parameters are $(s, q, \alpha)$, which represent the separation and mass 
ratio between the lens components, $M_1$ and $M_2$, and the angle between the source motion and 
the binary lens axis (source trajectory angle), respectively. For a 1L2S model, the extra parameters 
are $(t_{0,2}, u_{0,2}, q_F)$, which represent the time and separation at the closest approach of the 
source companion to the lens, and the flux ratio between the primary and secondary source components, 
$S_1$ and $S_2$. In the 1L2S modeling, we designate the parameters related to $S_1$ as 
$(t_{0,1}, u_{0,1})$ to distinguish them from those related to $S_2$. For all tested models, we 
additionally include the parameter $\rho$, which represents the ratio of the angular source radius 
$\theta_*$ to the Einstein radius $\thetae$, that is, $\rho=\theta_*/\thetae$.  This parameter 
(normalized source radius) is needed to account for finite-source effects that affect lensing light 
curves during the source passage over the lens or caustics.

In the modeling, we search for the set of lensing parameters (solution) that best describes the
observed data. The 2L1S solution is investigated in two steps, in which we search for the binary 
parameters $(s, q)$ using a grid approach in the first step, and then refine the individual local 
solutions found from the first step by allowing all parameters to vary in the second step. For the 
searches of the parameters other than the grid parameters, we use a downhill approach. We adopt 
this two-step procedure to identify possible degenerate solutions, in which different combinations 
of lensing parameters result in similar lensing light curves.  In the 1L2S modeling, we search for 
the solution using a downhill approach with initial values of $(t_{0,2}, u_{0,2}, q_F)$ assigned 
considering the time and strength of the anomaly. For the downhill approach, we use the Markov 
Chain Monte Carlo (MCMC) algorithm.

\begin{table}[t]
\small
\caption{Close~I 2L1S model parameters\label{table:two}}
\begin{tabular*}{\columnwidth}{@{\extracolsep{\fill}}lll}
\hline\hline
\multicolumn{1}{c}{Parameter}    &
\multicolumn{1}{c}{Standard}     &
\multicolumn{1}{c}{Higher-order} \\
\hline
$\chi^2$                     &   1637.2                  &  1601.9                   \\    
$t_0$ (HJD$^\prime$)         &   9417.020 $\pm$ 0.002    &  9417.020 $\pm$ 0.003     \\
$u_0$ (10$^{-2}$)            &   0.60 $\pm$ 0.02         &  0.62 $\pm$ 0.02          \\
$\te$ (days)                 &   69.62 $\pm$ 1.59        &  69.53 $\pm$ 2.31         \\
$s$                          &   0.920 $\pm$ 0.002       &  0.882 $\pm$ 0.015        \\
$q$ (10$^{-5}$)              &   0.84 $\pm$ 0.17         &  2.77 $\pm$ 0.67          \\
$\alpha$ (rad)               &   0.009 $\pm$ 0.003       &  0.083 $\pm$ 0.012        \\
$\rho$ (10$^{-3}$)           &   2.47 $\pm$ 0.37         &  3.85 $\pm$ 0.62          \\
$\pi_{{\rm E},N}$            &                           &  0.58 $\pm$ 0.15          \\
$\pi_{{\rm E},E}$            &                           &  -0.11 $\pm$ 0.06         \\
$ds/dt$ (yr$^{-1}$)          &                           &  1.19 $\pm$ 0.50          \\
$d\alpha/dt$ (yr$^{-1}$)     &                           &  -1.76 $\pm$ 0.37         \\
\hline                                   
\end{tabular*}
\tablefoot{ ${\rm HJD}^\prime\equiv {\rm HJD}-2450000$.  
}
\end{table}

\begin{table}[t]
\small
\caption{Close~II 2L1S model parameters\label{table:three}}
\begin{tabular*}{\columnwidth}{@{\extracolsep{\fill}}lll}
\hline\hline
\multicolumn{1}{c}{Parameter}     &
\multicolumn{1}{c}{Standard}      &
\multicolumn{1}{c}{Higher-order}  \\
\hline
$\chi^2$                     &  1626.9                   &  1592.4                  \\    
$t_0$ (HJD$^\prime$)         &  9417.023 $\pm$ 0.002     &  9417.025 $\pm$ 0.002    \\
$u_0$ (10$^{-2}$)            &  0.63 $\pm$ 0.02          &  0.62 $\pm$ 0.03         \\
$\te$ (days)                 &  67.38 $\pm$ 1.45         &  68.00 $\pm$ 2.67        \\
$s$                          &  0.918 $\pm$ 0.002        &  0.914 $\pm$ 0.014       \\
$q$ (10$^{-5}$)              &  1.21 $\pm$ 0.20          &  1.05 $\pm$ 0.46         \\
$\alpha$ (rad)               &  0.072 $\pm$ 0.004        &  0.104 $\pm$ 0.008       \\
$\rho$ (10$^{-3}$)           &  3.41 $\pm$ 0.51          &  2.68 $\pm$ 0.63         \\
$\pi_{{\rm E},N}$            &                           &  0.15 $\pm$ 0.28         \\
$\pi_{{\rm E},E}$            &                           &  0.00 $\pm$ 0.09         \\
$ds/dt$ (yr$^{-1}$)          &                           &  0.13 $\pm$ 0.46         \\
$d\alpha/dt$ (yr$^{-1}$)     &                           &  -0.87 $\pm$ 0.65        \\
\hline                                   
\end{tabular*}
\end{table}

\begin{table}[h]
\small
\caption{Wide 2L1S model parameters\label{table:four}}
\begin{tabular*}{\columnwidth}{@{\extracolsep{\fill}}lll}
\hline\hline
\multicolumn{1}{c}{Parameter}     &
\multicolumn{1}{c}{Standard}      &
\multicolumn{1}{c}{Higher-order}  \\
\hline
$\chi^2$                     &  1623.8                   &  1611.6                   \\    
$t_0$ (HJD$^\prime$)         &  9417.023 $\pm$ 0.002     &  9417.027 $\pm$ 0.003     \\     
$u_0$ (10$^{-2}$)            &  0.59 $\pm$ 0.02          &  0.60 $\pm$ 0.02          \\     
$\te$ (days)                 &  68.41 $\pm$ 1.55         &  67.78 $\pm$ 1.59         \\     
$s$                          &  1.089 $\pm$ 0.002        &  1.109 $\pm$ 0.020        \\     
$q$ (10$^{-5}$)              &  0.12 $\pm$ 0.04          &  0.17 $\pm$ 0.40          \\     
$\alpha$ (rad)               &  3.177 $\pm$ 0.001        &  3.177 $\pm$ 0.032        \\     
$\rho$ (10$^{-3}$)           &  0.88 $\pm$ 0.21          &  1.01 $\pm$ 0.09          \\     
$\pi_{{\rm E},N}$            &                           &  -0.06 $\pm$ 0.29         \\     
$\pi_{{\rm E},E}$            &                           &  0.11 $\pm$ 0.08          \\     
$ds/dt$ (yr$^{-1}$)          &                           &  -0.56 $\pm$ 0.63         \\     
$d\alpha/dt$ (yr$^{-1}$)     &                           &  -0.15 $\pm$ 0.75         \\     
\hline                                   
\end{tabular*}
\end{table}

\subsection{Binary-lens (2L1S) interpretation}\label{sec:three-one}

According the planetary 2L1S interpretation, the planet-host separation $s$ can be heuristically 
estimated from the location of the anomaly in the lensing light curve. Under the approximation of 
a very low-mass companion, the planetary caustic induced by a planet lies at the location with a 
separation from the primary lens of
\begin{equation}
u_{\rm a}\sim s-{1\over s}, 
\label{eq1}
\end{equation}
which has a negative value for a close binary with $s<1.0$ and a positive value for a wide binary
with $s>1.0$ \citep{Griest1998, Han2006}. Then, the planet-host separation is obtained by
solving Equation~(\ref{eq1}) with respect to $s$, that is,
\begin{equation}
s= {1\over 2} \left[ u_{\rm a} + \left( u_{\rm a}^2 + 4 \right)^{1/2} \right].
\label{eq2}
\end{equation}
The $u_{\rm a}$ value corresponds to the source separation (from $M_1$) at the time of the anomaly, 
$t_{\rm a}$, and it is related to the lensing parameters by
\begin{equation}
u_{\rm a} = \left[ \left({t_{\rm a} - t_0 \over \te} \right)^2 + u_0^2\right]^{1/2}.
\label{eq3}
\end{equation}
With the time of the anomaly at $t_{\rm a}\sim 9428.5$ and the 1L1S lensing parameters, $(t_0, u_0, 
\te)\sim (9417.02, 6.0\times 10^{-3}, 70$~days), the source separation is $u_{\rm a}\sim \pm 0.16$. 
Then, there exist two solutions of the planet separations of
\begin{equation}
s_c \sim 0.92\;\qquad  {\rm and} \qquad  s_w\sim 1.09,
\label{eq4}
\end{equation}
where the former and latter values correspond to the close and wide solutions, respectively.

\begin{figure*}[t]
\centering
\includegraphics[width=12.5cm]{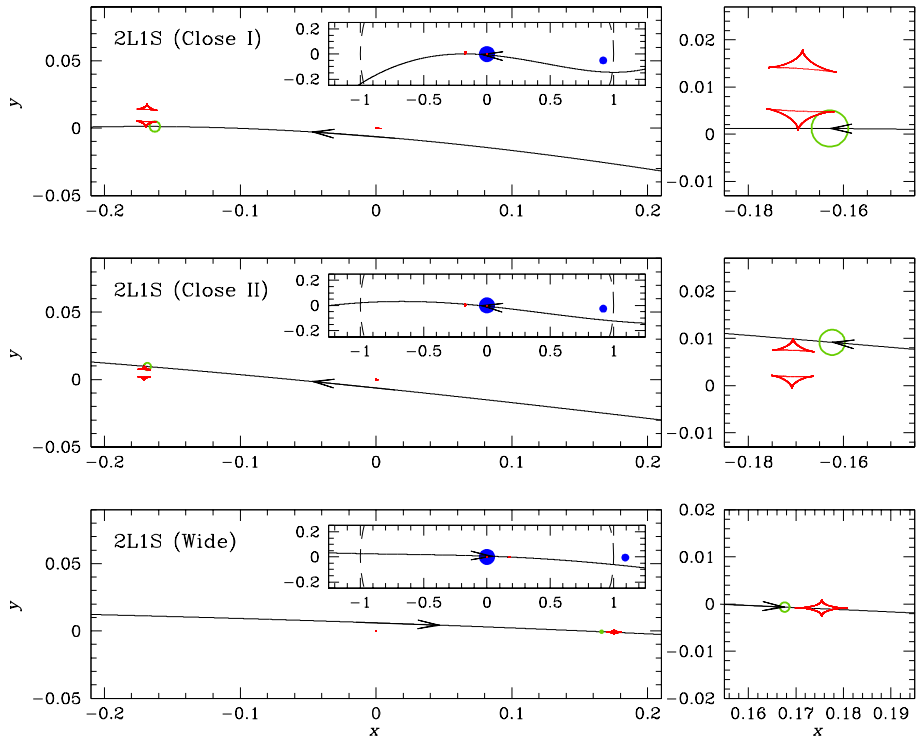}
\caption{
Lens system configurations of the three 2L1S solutions: close~I, close~II, wide solutions. In each 
panel, the small figures composed of concave curves represent the caustics and the curve with an 
arrow denotes the source trajectory. For each solution, the inset of the left panel shows the whole 
view including the lens positions (marked by blue dots) and the Einstein ring (dashed circle). The 
panel on the right side shows the zoom-in view around the planetary caustic.  The green empty circle 
on the source trajectory represents the source for which the size is scaled to the caustic size.  
Lengths are scaled to the Einstein radius and the coordinates are centered at the position of the 
primary of the planetary system.  The lensing parameters corresponding to the individual solutions 
are listed in Tables~\ref{table:two} -- \ref{table:four}.  
}
\label{fig:four}
\end{figure*}

From the 2L1S modeling, we identify 3 local solutions. Figure~\ref{fig:two} shows the locations 
of these local solutions in the $\chi^2$ map on the $\log s$--$\log q$ (left panel) and 
$\log s$--$\alpha$ (right panel) planes. The binary lensing parameters of the individual solutions 
are $(\log s, \log q, \alpha)\sim (-0.04, -5.1, 0.01)$, $\sim (-0.04, -5.1, 0.07)$, and 
$\sim (0.04, -5.9, 3.18)$ in logarithmic scale and radians, or $(s, q, \alpha)\sim (0.9, 0.9\times 
10^{-5}, 0.6^\circ)$, $\sim (0.9, 0.9\times 10^{-5}, 4.0^\circ)$, and $\sim (1.1, 0.13\times 10^{-5}, 
182.2^\circ)$ in linear scale and degrees. We refer to the individual solutions as ``close~I'', 
``close~II'', and ``wide'' solutions, respectively.  We note that the $s$ and $q$ values for the two 
close solutions are similar to each other, but they have slightly different values of the source 
trajectory angle, as shown in the inset of the right panel. The binary separations of the close, 
$s_c$, and wide solution, $s_w$, match well the heuristically estimated values in Equation~(\ref{eq4}). 
We note that although the binary separations $s_c$ and $s_w$ are approximately in the relation of 
$s_c\sim 1/s_w$, the degeneracy between the close and wide solutions has a different origin from the 
close--wide degeneracy between a pair of solutions for central planet-induced anomalies appearing in 
high-magnification events \citep{Griest1998, Dominik1999}. We will discuss in detail the origins 
of the degeneracies below.  Regardless of the solutions, the estimated mass ratios are very small,  
of the order of $10^{-5}$ for the close solutions or even smaller for the wide solutions, indicating 
that the companion to the lens is a planetary-mass object with a very small planet-to-host mass ratio 
according to the 2L1S interpretation.

The model curves and residuals of the individual 2L1S solutions around the region of the anomaly
are shown in Figure~\ref{fig:three}, and their lensing parameters are listed in Tables~\ref{table:two} 
-- \ref{table:four} together with the $\chi^2$ values of the fits. For each solution, we present two 
sets of solutions, in which one solution is obtained under the assumption of a rectilinear relative 
lens-source motion, standard model, and the other solution is obtained by considering the deviation 
from the rectilinear motion caused by the orbital motions of Earth \citep[microlens-parallax 
effect:][]{Gould1992a} and the binary lens \citep[lens-orbital effect:][]{Dominik1998}, higher-order 
model.  We will further discuss these higher-order effects below. From the comparison of the solutions, 
it is found that the close~II solution provides the best fit to the data, but the $\chi^2$ differences 
from the other solutions, $\Delta\chi^2=9.5$ with respect to the close~I solution and 19.2 with respect 
to the wide solution, are relatively modest. As a rule of thumb, differences of $\Delta\chi^2>10$ are 
required to distinguish between solutions with reasonable confidence, but careful consideration must 
be given when the differences are not much larger, as in the present case.

Figure~\ref{fig:four} shows the configurations of the lens systems for the three degenerate 2L1S 
solutions. In each panel, the small closed figures composed of concave curves represent the caustics 
induced by the planet, and the curve with an arrow represents the source trajectory. The configurations 
correspond to the models obtained with the consideration of the higher-order effects, and thus the 
source trajectory is slightly curved due to the microlens-parallax effect. The caustic varies in time due 
to the lens orbital motion, and the presented caustics correspond to the ones at the time of the anomaly. 
The planetary companion induces two sets of caustics, in which one is located very close to the host 
of the planet (central caustic) and the other is located away from the host (planetary caustic).  The 
planetary caustic induced by a close planet is located on the opposite side of the planet with respect 
to the host and is composed of two sets of closed curves, while the planetary caustic induced by a wide 
planet is located on the planet side and is composed of a single closed curve \citep{Han2006}. 
According to the close~I 
and II solutions, the anomaly was produced by the source passage over the upper and lower planetary 
caustics induced by a close planet, respectively. According to the wide solution, on the other hand, 
the anomaly was generated by the source passage over the planetary caustic of a wide planet. According 
to all 2L1S solutions, the source did not cross the central caustic. As a result, the central region 
does not exhibit an obvious anomalous feature, but we will discuss possible central anomalies in 
Sect.~\ref{sec:three-three}.

The form of the degeneracy among the three 2L1S solutions is very rare because it arises from the 
special configuration of the lens system.  The degeneracy between the close~I and II solutions, which 
we refer to as the ``upper-lower degeneracy'', occurs due to the special lens-system configuration, in 
which the source moves almost in parallel with the planet-host axis and passes through the upper and 
lower parts of the planetary caustic with similar patterns of magnification excess. The degeneracy between 
the close and wide solutions, which we referred to as the ``close-wide degeneracy for planetary caustics'', 
arises due to the small caustic size caused by the very low planet/host mass ratio.  In this case, it is 
difficult to delineate the detailed structure of the anomaly not only because of the short duration caused 
by the low mass ratio but also because of the featureless structure caused by finite-source effects.  This 
degeneracy is different from the close-wide degeneracy for central anomalies, arising from the intrinsic 
similarity between the central caustics induced by a close and a wide planet, in the sense that the anomaly 
was produced by planetary caustics.  The planetary caustics induced by a close and a wide planet differ both 
in shape and number, resulting in different patterns of magnification excess
\citep{Chung2005}.
Then, the close-wide degeneracy 
for planetary caustics is an accidental degeneracy arising from the inadequate precision and cadence of 
observations for very short anomalies induced by planets with very low mass ratios.

\begin{figure}[t]
\includegraphics[width=\columnwidth]{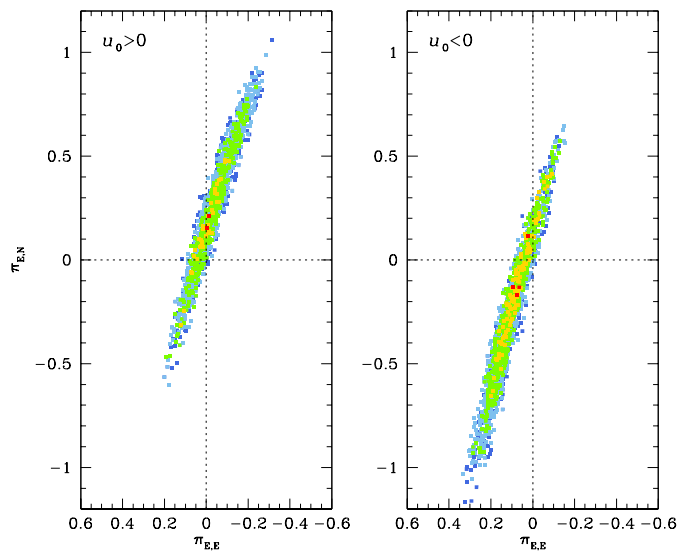}
\caption{
Scatter plot of MCMC points on the $\pi_{{\rm E},N}$--$\pi_{{\rm E},E}$ parameter plane. Color
coding
is same as that in Fig.~\ref{fig:two}, except that $n=1$.
The right and left panels are the plots for the two degenerate solutions with 
$u_0>0$ and $u_0<0$ solutions, respectively.
}
\label{fig:five}
\end{figure}

We check the feasibility of measuring the additional observables of the angular Einstein radius 
$\thetae$ and microlens parallax $\pie$, which can constrain the mass and distance to the lens.  The 
prerequisite for the estimation of $\thetae$ is the measurement of the normalized source radius $\rho$, 
because the Einstein radius is estimated by
\begin{equation}
\thetae = {\theta_* \over \rho},
\label{eq5}
\end{equation}
where the angular source radius $\theta_*$ can be estimated from the color and brightness of the 
source.  We will discuss the detailed procedure of $\theta_*$ measurement in Sect.~\ref{sec:four}. 
The $\rho$ value is measured by analyzing the planetary anomaly that is affected by finite-source 
effects, and the values corresponding to the individual 2L1S solutions are listed in Tables~\ref{table:two} 
-- \ref{table:four}.  We note that the $\rho$ values vary in the range of [0.88 -- 3.41]$\times 10^{-3}$ 
depending on the solutions.

\begin{figure}[t]
\includegraphics[width=\columnwidth]{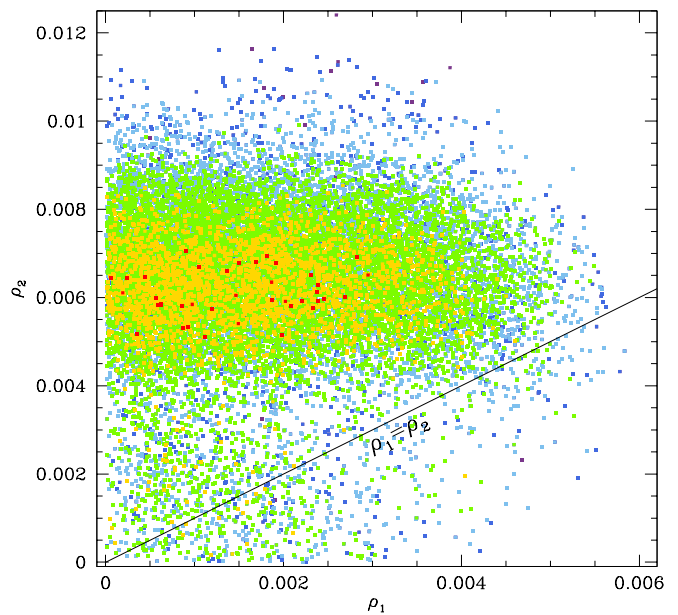}
\caption{
Scatter plot of points in the MCMC chain on the $\rho_1$--$\rho_2$ plane for the 1L2S model. 
Color coding is the same as that in Fig.~\ref{fig:five}.  The black line represents the equation 
$\rho_1 = \rho_2$.
}
\label{fig:six}
\end{figure}

The $\pie$ value is measured from the deviation in the light curve caused by the microlens-parallax 
effect.  For the measurement of $\pie$, we conduct additional sets of modeling for the individual 2L1S 
solutions by considering the parallax effect. Because it is known that the orbital motion of the 
lens can result in similar deviation to that induced by the parallax effect \citep{Batista2011}, we 
simultaneously consider both the microlens-parallax and lens-orbital effects \citep{Dominik1998} in 
the modeling. For the consideration of these higher-order effects, we add four extra parameters in 
the modeling: $(\pi_{{\rm E},N}, \pi_{{\rm E},E})$ for the microlens parallax effect and $(ds/dt, 
d\alpha/dt)$ for the lens-orbital effect. The two parallax parameters denote the north and east 
components of the microlens-parallax vector $\pivec_{\rm E}=(\pi_{\rm rel}/\thetae)(\muvec/\mu)$, 
respectively, and the two orbital parameters denote the annual change rates of the binary separation 
and source trajectory angle, respectively

The lensing parameters of the individual 2L1S solutions considering the higher-order effects are 
listed in Tables~\ref{table:two} -- \ref{table:four}.  It is found that the fits of the higher-order 
models improve by $\Delta\chi^2=35.2$, 34.5, and 12.2 with respect to the standard models for the close~I, 
close~II, and wide solutions, respectively.  We note that the uncertainties of the planet parameters $(s, q)$ 
are bigger than those estimated from the standard models, although the fits of the higher-order models improve.  
This is a general tendency in lens modeling arising because the relative motion of the source with respect to 
the caustic has more degree of freedom.  Figure~\ref{fig:five} shows the scatter plot of points in the MCMC 
chain on the $\pi_{{\rm E},E}$--$\pi_{{\rm E},N}$ parameter plane. The plot is based on the close~II model, 
which yields the best fit among the solutions, and the other models result in similar scatter plots. It 
shows that the component of $\pivec_{\rm E}$ that is parallel with the direction of Earth's acceleration, 
$\pi_{{\rm E},\parallel}$, is well constrained, but the uncertainty of the perpendicular component, 
$\pi_{{\rm E},\perp}$, is considerable.  We also check the so-called ecliptic degeneracy between a pair 
of solutions with $u_0>0$ and $u_0<0$ arising due to the mirror symmetry of the source trajectory with 
respect to the binary axis \citep{Skowron2011}.  The scatter plot for the $u_0<0$ solution is also presented 
in Figure~\ref{fig:five}.  It is found that the $u_0<0$ solution results in a similar pattern to that of 
the $u_0>0$ solution.

\subsection{Binary-source (1L2S) interpretation}\label{sec:three-two}

We also check the interpretation in which the observed anomaly was produced by a companion
to the source. For a lensing event involved with two source stars, the observed flux, $F$, is 
the superposition of fluxes from the events associated with the individual source stars
\citep{Griest1992, Han1997}, that is,
\begin{equation}
F = F_1(A_1 + q_F A_2). 
\label{eq6}
\end{equation}
Here $F_1$ represents the flux of the primary source, and $A_1$ and $A_2$ denote the lensing
magnifications associated with the individual source stars, that is,
\begin{equation}
A_i = {u_i^2+2 \over u_i (u_i^2+4)^{1/2} };\qquad
u_i = \left[ u_{0,i}^2+ \left( {t-t_{0,i} \over \te}\right)^2\right]^{1/2}.
\label{eq7}
\end{equation}
In Table~\ref{table:five}, we list the lensing parameters obtained from the 1L2S modeling, and 
the model curve and residual around the anomaly are shown in Figure~\ref{fig:three}.  The values 
$\rho_1$ and $\rho_2$ denote the normalized radii of the primary and companion source stars, 
respectively.  It is found that the 1L2S model also well describes the anomaly, yielding a fit 
that is comparable to those of the 2L1S models.

\begin{table}[t]
\small
\caption{1L2S model parameters\label{table:five}}
\begin{tabular*}{\columnwidth}{@{\extracolsep{\fill}}lll}
\hline\hline
\multicolumn{1}{c}{Parameter}    &
\multicolumn{1}{c}{Value}    \\
\hline
$\chi^2$                    &   1626.6                  \\    
$t_{0,1}$ (HJD$^\prime$)    &   9417.022 $\pm$ 0.002    \\
$u_{0,1}$ (10$^{-2}$)       &   0.60 $\pm$ 0.02         \\
$t_{0,2}$ (HJD$^\prime$)    &   9428.575 $\pm$ 0.067    \\
$u_{0,2}$ (10$^{-2}$)       &   0.22 $\pm$ 0.13         \\
$\te$ (days)                &   67.91 $\pm$ 1.70        \\
$\rho_1$ (10$^{-3}$)        &   $< 4.0$                 \\
$\rho_2$ (10$^{-3}$)        &   6.41 $\pm$ 2.32         \\
$q_F$                       &   0.003 $\pm$ 0.001       \\
\hline                                   
\end{tabular*}
\end{table}

Although the 1L2S model well describes the observed anomaly, it is rejected due to its unphysical 
lensing parameters. According to the model, the flux from the source companion comprises a very 
minor fraction, $q_F\sim 0.3\%$, of the flux from the primary source, but the estimated normalized 
radius of the source companion, $\rho_2\sim 6.4\times 10^{-3}$, is significantly greater than the 
upper limit of the primary source, $\rho_{1,{\rm max}}\sim 4.0\times 10^{-3}$. This is shown in 
Figure~\ref{fig:six}, in which we plot the points in the MCMC chain on the $\rho_1$--$\rho_2$ 
plane. With the contradiction that the primary source is smaller than its very faint companion, 
we rule out the 1L2S interpretation of the event.

\begin{figure}[t]
\includegraphics[width=\columnwidth]{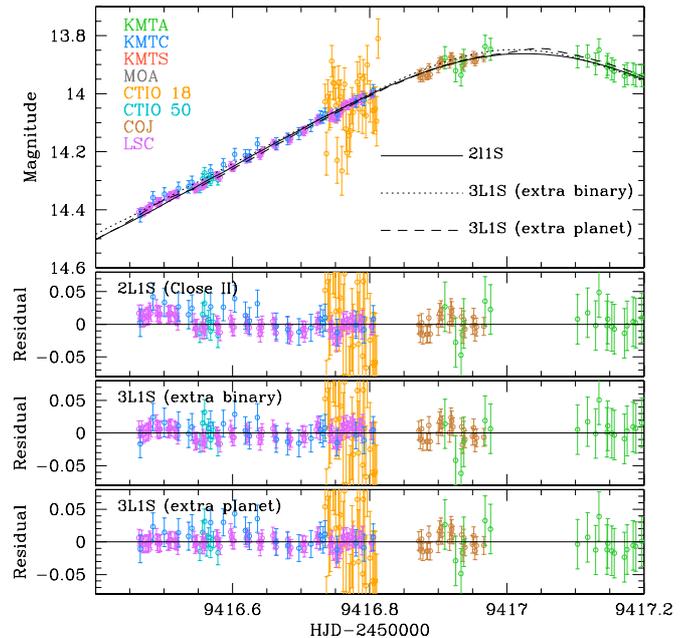}
\caption{
Zoom-in view of the peak region.  The curves drawn over the data points are the 2L1S model (close~II) 
and two 3L1S models. The lower panels shows the residuals from the individual models.
}
\label{fig:seven}
\end{figure}

\subsection{Extra lens component: 3L1S interpretation}\label{sec:three-three}

Although the peak region of the light curve does not exhibit an obvious anomalous feature, the
region is susceptible to deviations induced by an extra lens component. This is because the event
reached a very high magnification, implying that the source passed through the central magnification 
region, around which an extra lens component, if it exists, would induce a caustic affecting the 
magnification pattern of the region \citep{Gaudi1998b}.  We recall from Sect.~\ref{sec:two} that 
the peak was subjected to high-cadence followup and auto-followup observations, which significantly 
enhance the sensitivity to low-amplitude anomalies.  Therefore, we inspect small deviations in the 
peak region.

The peak region of the light curve is shown in top panel of Figure~\ref{fig:seven}. The solid curve 
drawn over the data points is the model curve of the 2L1S (close~II) solution, and the residual from 
the model is shown in the second panel. It is found that the model leaves small deviations with 
$\lesssim 0.03$~mag level mainly in the LSC data set taken during $9416.46\lesssim {\rm HJD}^\prime
\lesssim 9416.80$.  Although the deviation is subtle, we check the possibility of a third lens 
component, $M_3$, for the origin of the deviation.

The lensing modeling with three lens components (3L1S model) requires one to include extra parameters 
in addition to those of the 2L1S modeling. These parameters are $(s_3, q_3, \psi)$, which represent 
the separation and mass ratio between $M_1$ and $M_3$, and the orientation angle of $M_3$ with respect 
to the $M_1$-$M_2$ axis with a center at the position of $M_1$, respectively.  The 3L1S modeling is done 
in two steps, in which we conduct grid searches for the parameters related to $M_3$, that is, $(s_3, 
q_3, \psi)$, in the first step, and then refine the local solutions found from the grid search by 
allowing all parameters to vary. In the grid search, we set the other parameters by adopting those 
of the 2L1S solution (close~II solution) under the assumption that the anomaly induced by a tertiary 
lens component would be confined in a small region around the peak, and thus could be treated as a 
perturbation \citep{Bozza1999, Han2001}.

From the 3L1S modeling, we find two sets of local solutions. According to one set, the peak
residual is reduced by a low-mass binary companion with a projected separation either much
bigger or smaller than the Einstein radius (``extra-binary'' solution), and according to the other
solution, the residual is explained by a second planet with a separation similar to the Einstein
radius (``extra-planet'' solution). The triple-lens parameters of the solutions are
\begin{equation}
(\log s_3, \log q_3, \psi) \sim 
\begin{cases}
(\pm 0.68, -0.82, 1.67),    & \text{extra binary},\\
(    0.0,  -5.21, 2.01),    & \text{extra planet}.
\end{cases}
\label{eq8}
\end{equation}

\begin{figure}[t]
\includegraphics[width=\columnwidth]{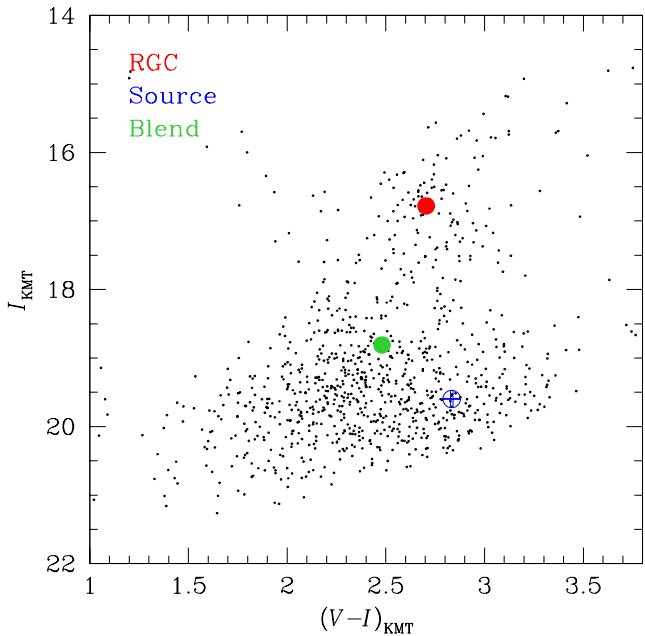}
\caption{
Locations of the source, blend, and RGC centroid in the instrumental CMD of stars around
the source. 
}
\label{fig:eight}
\end{figure}

Figure~\ref{fig:seven} shows the models curves and residuals of the extra-binary (for the wide 
solution with $s\sim 4.8$) and extra-planet 3L1S solutions. Although the improvement of the fit in the 
peak region supports the possibility of an extra lens component, it is difficult to claim a secure 
detection of a tertiary lens component due to several reasons. First, the fit improvement is marginal:
 $\Delta\chi^2\sim 52.2$ and 34.6 with respect to the 2L1S solution for the extra binary and planet 
solutions, respectively.  Second, the two data sets covering the peak anomaly, LSC and KMTC data sets, 
do not show a strong consistency as shown in the bottom two panels of Figure~\ref{fig:seven}. Third, 
there exist multiple solutions, that is, extra-binary and extra-planet solutions, that can explain 
the central residual, making it difficult to uniquely specify the nature of the extra lens component 
even if the signal is real.

\section{Source star and Einstein radius}\label{sec:four}

In this section, we estimate the angular source radius in order to estimate the angular Einstein 
radius using the relation in Equation~(\ref{eq5}). We deduce $\theta_*$ from the color and magnitude 
of the source star. For the measurement of the reddening and extinction-corrected (dereddened) color 
and magnitude, $(V-I,I)_0$, from uncalibrated values, $(V-I, I)$, in the instrumental color-magnitude 
diagram (CMD), we use the \citet{Yoo2004} method, which utilizes the centroid of red giant clump (RGC), 
for which its dereddened color and magnitude, $(V-I, I)_{{\rm RGC},0}$, are known, for calibration.

Figure~\ref{fig:eight} shows the locations of the source and RGC centroid in the instrumental CMD 
of stars around the source constructed using the pyDIA photometry \citep{Albrow2017} of the KMTC data. 
The values of $(V-I, I)$ are measured from regression of the $I$ and $V$-band KMTC data processed 
using the pyDIA code with the change of the event magnification. Also marked is the location of 
the blend with $(V-I, I)_b=(2.48, 18.81)$. The measured instrumental color and magnitude are 
$(V-I, I)_s=(2.833\pm 0.006, 19.597\pm 0.001)$ for the source and $(V-I, I)_{\rm RGC}=(2.704, 16.780)$ 
for the RGC centroid. With the offsets in color and magnitude, $\Delta (V-I, I)$, between the source 
and RGC centroid together with the known dereddened color and magnitude of the RGC centroid, 
$(V-I, I)_{{\rm RGC},0}= (1.060, 14.510)$ \citep{Bensby2013, Nataf2013}, the dereddened values of 
the source are estimated as
\begin{equation}
\eqalign{
(V-I, I)_{s,0} 
& =  (V-I, I)_{{\rm RGC},0} + \Delta (V-I, I) \cr 
               & = (1.189\pm 0.006, 17.327\pm 0.001),        \cr
}
\label{eq9}
\end{equation}
indicating that the source is a bulge subgiant with a K4 spectral type. We convert the $V-I$ color
into $V-K$ color using the relation of \citet{Bessell1988}, and then interpolate the angular
source radius from the $(V-K)$--$\theta_*$ relation of \citet{Kervella2004}.  The estimated source 
radius from this procedure is
\begin{equation}
\theta_* = (1.87 \pm  0.13)~\mu{\rm as}.
\label{eq10}
\end{equation}
With the measured value of $\theta_*$, the angular Einstein radius is estimated by Equation~(\ref{eq5}) 
and the relative lens-source proper motion is computed by $\mu=\thetae/\te$. In Table~\ref{table:six}, 
we list the estimated values of $\thetae$ and $\mu$ corresponding to the three degenerate 2L1S solutions. 
We note that the value of the angular source radius does not vary depending on the solutions, but the 
values of $\thetae$ and $\mu$  have different values depending on the solutions because the individual 
solutions have different values of $\rho$.

\begin{table}[t]
\small
\caption{Angular Einstein radius and relative lens-source proper motion\label{table:six}}
\begin{tabular*}{\columnwidth}{@{\extracolsep{\fill}}lll}
\hline\hline
\multicolumn{1}{c}{Model }                  &
\multicolumn{1}{c}{$\thetae$ (mas)}         &
\multicolumn{1}{c}{$\mu$ (mas yr$^{-1}$)}   \\
\hline
Close~I     &  0.49 $\pm$ 0.09   &  2.57 $\pm$ 0.478  \\
Close~II    &  0.70 $\pm$ 0.18   &  3.76 $\pm$ 0.978  \\
Wide        &  1.85 $\pm$ 0.20   &  9.96 $\pm$ 1.11   \\
\hline                                   
\end{tabular*}
\end{table}

We check whether a significant fraction of blended flux comes from the lens by measuring the astrometric 
offset between the image centroid at the baseline and the source position measured from the difference 
image.  For the measurement of the source position, we use an image taken with the use of the 3.6~m 
Canada-France-Hawaii Telescope under a good seeing ($0^{\prime\prime}\hskip-2pt .65$) condition at the 
time around the peak.  The measured offset is
\begin{equation}
\Delta\thetavec (N, E) = (26\pm 17, -86\pm 8)~{\rm mas}.
\label{eq11}
\end{equation}
The offset is much bigger than the uncertainty, indicating that the major origin of the blended flux 
is not the lens. It must therefore be either a companion to the source, a companion to the lens, or an 
ambient star. Note that if the blended light is due to a star in or near the bulge, then its color and 
absolute magnitude $[(V-I)_0,M_I]\simeq (0.86,2.0)$ indicate that it is a subgiant. Based on the surface 
density of subgiants toward this field, the chance that an ambient subgiant would be projected within 
$\sim 100$~mas of the lens is about $p\sim 6\times 10^{-4}$.  In addition, the probability that the 
source would have a subgiant companion at projected separation $\gtrsim 700$~AU is also relatively low, 
roughly $p\sim 3\times 10^{-3}$. A similar probability applies to companions to the lens. Thus, we cannot 
reliably distinguish among these possibilities based on only the measured astrometric offset.  The main 
impact of the blended light on the analysis is that it places an upper limit on the lens flux.

\begin{figure}[t]
\includegraphics[width=\columnwidth]{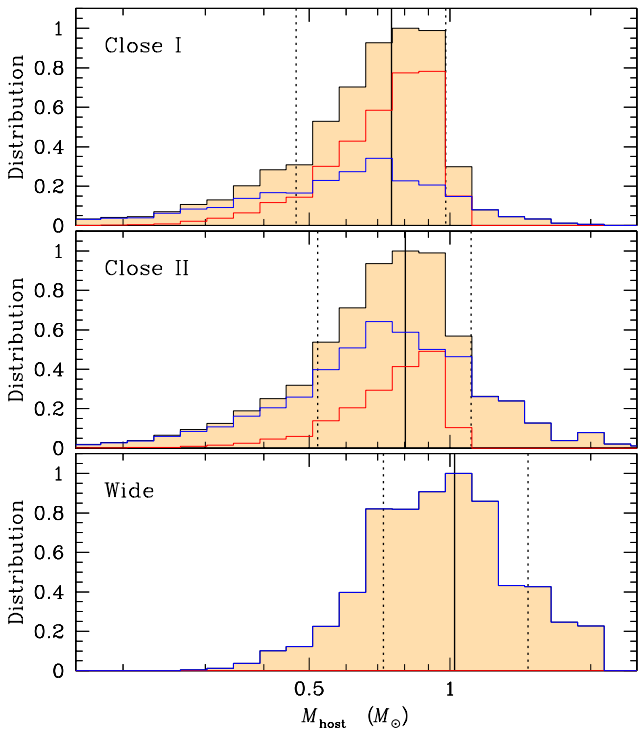}
\caption{
Posterior distributions of the lens mass constructed from the Bayesian analyses based on the 
observables of the three 2L1S solutions. For each distribution, the solid vertical line 
indicates the median, and the two dotted lines represent the 1$\sigma$ range of the distribution. 
The blue and red curves denote the contribution by the disk and bulge lens populations, and the 
solid distribution is the sum of the two populations. 
}
\label{fig:nine}
\end{figure}

\section{Physical parameters}\label{sec:five}

We estimate the physical parameters of the planetary system using the observables of the lensing
event. The mass and distance to the lens system can be uniquely determined with the observables
of $\pie$ and $\thetae$ using the relation
\begin{equation}
M = {\thetae \over \kappa\pie};\qquad
\dl = {{\rm AU} \over \pie\thetae + \pi_{\rm S}}.
\label{eq12}
\end{equation}
Here $\kappa =4G/(c^2{\rm AU})$ and $\pi_{\rm S}={\rm AU}/D_{\rm S}$ denotes the parallax of the 
source located at a distance $D_{\rm S}$ \citep{Gould2000}. For KMT-2021-BLG-0912, the 
uncertainty of $\pie$ is considerable, as shown in Figure~\ref{fig:five}, making it difficult to 
uniquely determine $M$ and $\dl$ using the relation in Equation~(\ref{eq12}), but the physical lens 
parameters can still be constrained with the remaining observables $\thetae$ and $\te$, which are 
related to the physical parameters by
\begin{equation}
\te = {\thetae\over \mu}; \qquad
\thetae = \left( \kappa M \pi_{\rm rel} \right)^{1/2},
\label{eq13}
\end{equation}
where $\pi_{\rm rel} = {\rm AU}( 1/D_{\rm L} - 1/D_{\rm S})$.  We estimate the physical parameters 
by conducting a Bayesian analysis with the constraints provided by $\thetae$ and $\te$ and using 
the Galactic model defining the physical and dynamical distributions and mass function of Galactic 
objects. Although $\pie$ is uncertain, we consider its constraint by imposing the restriction given 
by the covariance matrix of the parallax ellipse, that is, Figure~\ref{fig:five}. For the three 
degenerate solutions, the $\te$ values are similar to one another, but the $\thetae$ values vary 
substantially depending on the solutions, and thus we conduct three sets of analysis for the individual 
solutions.

\begin{figure}[t]
\includegraphics[width=\columnwidth]{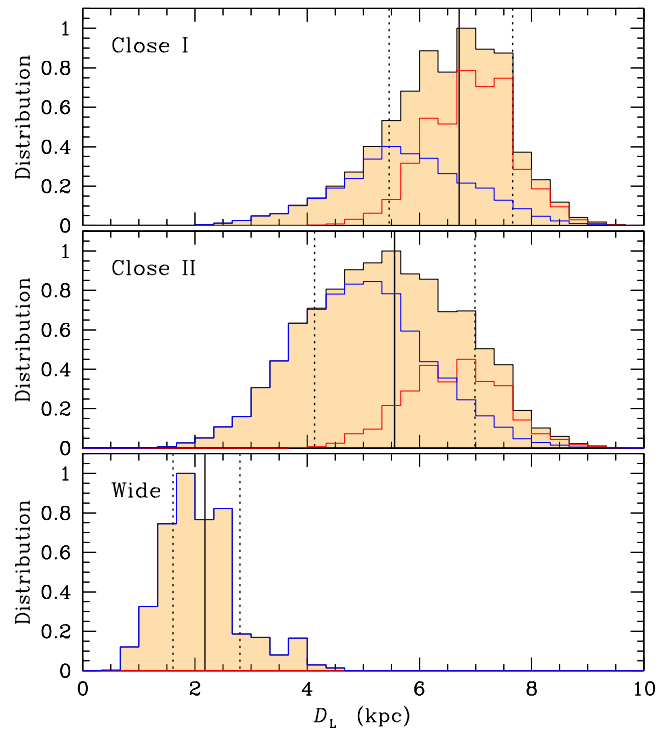}
\caption{
Posterior distributions of the distance to the lens. Notations are same as those in Fig.~\ref{fig:nine}.
}
\label{fig:ten}
\end{figure}

The Bayesian analysis is carried out by producing a large number ($6\times 10^6$) of artificial 
lensing events from a Monte Carlo simulation using the Galactic model. We adopt the \citet{Jung2021} 
Galactic model that is constructed based on the \citet{Robin2003} and \citet{Han2003} models for the 
physical distributions of disk and bulge objects, \citet{Jung2021} and \citet{Han1995} models for the 
dynamical distribution for disk and bulge objects, and the \citet{Jung2018} mass function.  In the mass 
function, we include brown-dwarf and white-dwarf populations of lenses, but exclude neutron stars and 
black holes because it is extremely unlikely that planets could remain bound after supernova explosions.  
For a more detailed description of the Galactic model, see \citet{Jung2021}.  With the artificial lensing 
events produced from the simulation, we construct the posterior distributions of $M$ and $\dl$ for the 
events with observables consistent with the measured values.

\begin{table}[t]
\small
\caption{Physical parameters\label{table:seven}}
\begin{tabular*}{\columnwidth}{@{\extracolsep{\fill}}llll}
\hline\hline
\multicolumn{1}{c}{Parameter}   &
\multicolumn{1}{c}{Close~I}     &
\multicolumn{1}{c}{Close~II}    &
\multicolumn{1}{c}{Wide}        \\
\hline
$M_{\rm host}$ ($M_\odot$)      &  $0.75^{+0.23}_{-0.28}$   & $0.80^{+0.31}_{-0.28}$  &  $1.02^{+0.45}_{-0.30}$   \\ [0.7ex]
$M_{\rm planet}$ ($M_\oplus$)   &  $6.92^{+2.13}_{-2.59}$   & $2.81^{+1.08}_{-0.99}$  &  $0.58^{+0.25}_{-0.17}$   \\ [0.7ex]
$\dl$ (kpc)                     &  $6.71^{+0.96}_{-1.25}$   & $5.56^{+1.43}_{-1.43}$  &  $2.18^{+0.63}_{-0.57}$   \\ [0.7ex]
$a_\perp$ (AU)                  &  $3.03^{+3.46}_{-2.46}$   & $3.14^{+3.94}_{-2.33}$  &  $3.81^{+4.89}_{-2.81}$   \\ [0.7ex]
%
\hline                                   
\end{tabular*}
\end{table}

Figures~\ref{fig:nine} and \ref{fig:ten} show the posterior distributions of $M_{\rm host}=M_1$, and 
$\dl$, respectively. In Table~\ref{table:seven}, we summarize the physical parameters of $M_{\rm host}$, 
$M_{\rm planet}=q M_{\rm host}$, $\dl$, and $a_\perp =s\dl \thetae$ estimated based on the three 2L1S 
solutions. For each parameter, the representative value is chosen as the median value of the Bayesian 
distribution, and the lower and upper limits are estimated as 16\% and 84\% of the distribution. 
The host is a K-type star according to the close solutions, while it is a G-type star according to the 
wide solution.  The estimated mass of the host for the wide solution is bigger than those expected from 
the close solutions because the input value of the Einstein radius ($\thetae\sim 1.85$~mas) for the wide 
solution is substantially bigger than those of the close solutions ($\thetae\sim 0.49$~mas for the close~I 
solution and $\sim 0.70$~mas for the close~II solution). For the same reason, the estimated 
distance to the lens is closer for the wide solution than the distances estimated from the close solutions. 
The mass of the planet estimated from the close solutions ($M_{\rm planet}\sim 6.9~M_\oplus$ for the close~I 
solution and $\sim 2.8~~M_\oplus$ for the close~II solution) are heavier than Earth, while the mass estimated 
from the wide solution ($M_{\rm planet}\sim 0.6~M_\oplus$) is below the mass of Earth.

Among the three 2L1S solutions, we exclude the wide solution for three major reasons.  First, the model 
fit of the wide solution is worse than the close~II solution by $\Delta\chi^2=19.2$ when the higher-order 
models are compared.  This $\chi^2$ difference is already quite large and significantly disfavors the wide 
solution by itself.  However, the wide solution yields the best fit when the standard models are compared, 
and thus this alone does not provide a crucial clue for disfavoring the wide solution.  Second, only 
a very small fraction of Bayesian output is consistent with the flux constraint.  According to the Bayesian 
analysis, the median values of the lens mass and distance are $(M_{\rm host},\dl)\sim (1~M_\odot, 2.2~{\rm kpc})$, 
and thus the lens would have a dereddened magnitude of $I_0\sim 15.8$.  By adopting $A_I \sim 1.5$ considering 
that most of the dust is very nearby because it is a high latitude field ($b \sim 4^\circ$), the apparent 
magnitude of the lens would be $I\sim 17.3$, which is substantially brighter than the blend with $I_b\sim 18.8$.  
Taking account of the fact that the lens mass could be more than $1\sigma$ lower than the median, the lower 
limit of the lens mass is $M_{\rm min}\sim 0.7~M_\odot$, and the corresponding lower limit of the lens magnitude 
is $I_{\rm L, min}\sim 19.0$.  Considering that the flux from the lens comprises a small fraction of the blending 
flux as discussed in Sect.~\ref{sec:four}, the lens should be much fainter than $I_b$.  This implies that only 
a very small fraction of Bayesian output is consistent with the constraint given by the blending flux.  Finally, 
the wide solution results in an overall Bayesian probability that is much lower than the other solutions.  The 
relative ratios of the overall Bayesian probabilities are $1.00:0.83:0.04$ for the close~I, close~II, and wide 
solutions, respectively.  This implies that the probability of a lensing event with $M$ and $\dl$ corresponding 
to the wide solution is much lower than the probability of an event with the physical lens parameters corresponding 
to the close solutions.

Despite the variation depending on the remaining solutions, the estimated mass indicates that
KMT-2021-BLG-0912Lb is a super Earth, which is defined as a planet having a mass higher than
that of Earth, but substantially lower than those of the Solar System's ice giants \citep{Valencia2007}. 
The number of super Earth planets detected by microlensing is rapidly increasing since the onset of 
high-cadence surveys.  The microlensing planets with $M \lesssim  10~M_\oplus$ are listed in Table~3 of 
\citet{Han2021}.  With an additional planet KMT-2020-BLG-0414LAbB \citep{Zang2021}, that has been detected 
since the publication of their paper, there exist 19 super-Earth microlensing planets with $M\lesssim 
10~M_\oplus$, including the one we report in this work. Among these planets, 16 planets were detected 
during the last 6 years after the full operation of the KMTNet survey. This well demonstrates the 
importance of high-cadence global surveys in detecting low-mass planets.

\section{Conclusion}\label{sec:six}

We investigated the light curve of the lensing event KMT-2021-BLG-0912, which exhibited a very short 
anomaly relative to a  single-lens single-source form.  Although this event reached a high magnification 
and had substantial followup data, the anomaly occurred $\sim 11.5$~days after the peak when the 
event was only magnified by $A\sim 6$ and no longer receiving followup observations. Hence, the 
discovery and characterization of this anomaly relied entirely on survey observations taken at a 
cadence of several points per night from multiple locations.

From the modeling of the light curve under various interpretations, we found four solutions, in 
which three solutions were found under the assumption that the lens was composed of two masses, 
and the other solution was found under the assumption that the source was comprised of a binary-star 
system.  The 1L2S model was rejected based on the contradiction that the faint source companion was 
bigger than its primary, and one of the 2L1S solutions was excluded from the combination of the 
relatively worse fit, blending constraint, and lower overall probability,  leaving two surviving 
solutions with planet/host mass ratios of $\sim 2.8 \times 10^{-5}$ and $\sim 1.1\times 10^{-5}$.  
A subtle central deviation might support the possibility of a tertiary lens component, either a 
binary companion to the host with a very large or small separation or a second planet lying at 
around the Einstein ring, but it was difficult to claim a secure detection due to the marginal fit 
improvement, lack of consistency among different data sets, and difficulty in uniquely specifying 
the nature of the tertiary component.  With the observables of the event, it was estimated that the 
masses of the planet and host were $\sim (6.9~M_\oplus, 0.75~M_\odot)$ according to one solution and 
$\sim (2.8~M_\oplus, 0.80~M_\odot)$ according to the other solution, indicating that the planet was 
a super Earth around a K-type star, regardless of the solutions.  The fact that 16, including the 
one reported in this work, out of 19 microlensing super-Earth planets with $M\lesssim 10~M_\oplus$ 
were detected during the last 6 years well demonstrates the importance of high-cadence surveys in 
detecting very low-mass planets.

\begin{acknowledgements}
Work by C.H. was supported by the grants  of National Research Foundation of Korea 
(2020R1A4A2002885 and 2019R1A2C2085965).
S.D. and Z.L. acknowledges the science research grants from the China Manned Space Project with 
NO.~CMS-CSST-2021-A11.
W.Zang, S.M. and H.Y. acknowledge support by the National Science Foundation of China (Grant
No. 11821303 and 11761131004). This research uses data obtained through the Telescope Access
Program (TAP), which has been funded by the TAP member institutes.
J.C.Y. acknowledges support from N.S.F Grant No. AST-2108414.
This research has made use of the KMTNet system operated by the Korea Astronomy and Space 
Science Institute (KASI) and the data were obtained at three host sites of CTIO in Chile, 
SAAO in South Africa, and SSO in Australia.
Work by I.K. was supported by JSPS KAKENHI Grant Number
20J20633. Work by D.P.B., A.B., and C.R. was supported by NASA through
grant NASA-80NSSC18K027. T.S. acknowledges the financial support from the
JSPS, JSPS23103002, JSPS24253004, and JSPS2624702. Work by N.K. is supported
by JSPS KAKENHI Grant Number JP18J00897. The MOA project is
supported by J SPS KAK-ENHI Grant Number JSPS24253004, JSPS26247023,
JSPS23340064, JSPS15H00781, JP16H06287, 17H02871, and 19KK0082.
\end{acknowledgements}

\end{document}